\documentclass[aps,prstab,twocolumn,groupedaddress,nofootinbib,showpacs]{revtex4}
\usepackage{graphicx}
\usepackage{amsmath}

\begin{document}

\title{Longitudinal dynamics with RF phase modulation in the LNLS electron storage ring}

\author{N. P. Abreu}
\email{natalia@lnls.br}

\affiliation{Instituto de F\'isica "Gleb Wataghin", Universidade Estadual de Campinas, C.P. 6165, 13083-970 Campinas, Brazil}
\affiliation{LNLS, Laborat\'orio Nacional de Luz S\'incrotron, Cx. Postal 6192, CEP 13083-970, Campinas, Brazil}

\author{R. H. A. Farias and P. F. Tavares }

\affiliation{LNLS, Laborat\'orio Nacional de Luz S\'incrotron, Cx. Postal 6192, CEP 13083-970, Campinas, Brazil}



\date{\today}


\begin{abstract}
In the Brazilian synchrotron light source, we observed that
modulating the phase of the accelerating fields at approximately twice the
synchrotron frequency suppressed remarkably well a longitudinal
coupled-bunch mode of the beam driven by a higher order mode (HOM) in one of the radiofrequency
(RF) cavities. In this work, we present the results of a set of systematic
measurements, in single and multibunch mode, aimed at
characterizing the effects of RF phase modulation on the beam. We compare
those experiments with the results of tracking simulations
and of a theoretical model in which Landau damping is the stabilizing mechanism that explains
the suppression of the longitudinal coupled bunch instability.
We also measure the frequency of the stable islands
created in longitudinal phase space by phase modulation and the longitudinal beam transfer function (BTF) as a
function of the modulation frequency and amplitude.
The experimental results are in good agreement with theoretical expectations.
\end{abstract}



\maketitle

\section{INTRODUCTION}

Studies related to the effects of phase or amplitude modulation of the accelerating RF
fields in circular accelerators date back to the early 1990's when a series of classical experiments were performed with the aim of
understanding basic aspects of non-linear longitudinal beam dynamics at low current in hadron machines \cite{a,b,Huang-4,c,Mod1-5}. Apart
from the fundamental interest of those experiments, related to the possibility of direct observations of non-linear
phenomena such as the formation of islands in phase space and measurement of synchrotron tune variations
with amplitude, the main applications envisaged by those works were related to issues such as lifetime limitations due to
rf noise, the development of super-slow extraction techniques and the use of parametric feedback for multibunch instabilities.
Those early works laid the basic theoretical foundations for subsequent interest in applying RF phase or voltage
modulation to electron storage rings used as synchrotron light sources \cite{e,g} as a tool to control electron bunch density \cite{f,Mod4-3},
increase beam lifetime \cite{Mod3-2} and reduce the amplitude or even suppress coupled-bunch instabilities \cite{Mod2-1,Mod3-2,d}, resulting in relevant improvement to the
performance of those machines. Some of those previous works analyzed the effects of RF phase modulation at the first and third harmonics of the
synchrotron frequency \cite{Mod3-2,Mod5,Orsini}, whereas others (such as the present paper) \cite{Mod2-1} focus on modulation at the second harmonic.

However, in spite of the success of the implementation of such techniques at several laboratories,
some aspects of the physical mechanisms that explain the effectiveness of the reduction of the amplitude of coupled-bunch modes
(CBMs) are still not completely understood and a quantitative theoretical description is lacking. There is also some difficulty in
simulating the effects of phase modulation in multibunch mode; most of the work done so far simulates a single bunch and it is not
obvious how to infer the multibunch behavior from these results.

All experiments reported here were conducted at the LNLS 1.37 GeV electron storage ring \cite{UVX}, which routinely operates to produce synchrotron radiation for users. The original motivation for our interest in RF phase modulation
was the need to circumvent longitudinal instabilities that were observed in the LNLS ring right after the installation
of a second RF cavity by the end of 2003 \cite{UpgradedLNLS}. This cavity has a longitudinal HOM (with a frequency of 903 MHz)
which caused an intermittent orbit distortion that mimics the second order
dispersion function and which appeared when the beam coupled with the HOM
of the cavity exciting a large longitudinal dipolar oscillation. With temperature and plunger scans we were
able to identify the longitudinal mode L1 (associated with the CBM
$\sharp133$) of the new RF cavity as the main source of instabilities in
the machine. Since it was not possible to find a cavity operational
condition that would be free from
instabilities, an active solution in the form of phase modulation
of the RF fields at approximately twice the synchrotron frequency was attempted
with success, having a noticeable impact on CBM
amplitudes and helping to alleviate the orbit fluctuation.

In this work, we develop a theoretical model and a simulation tool to analyze the effects of phase modulation on the longitudinal beam
dynamics. According to our model, RF phase modulation enhances
stability through the increase of the incoherent frequency spread
that results as the electrons populate the phase-space
islands created by the action of the modulation. In order to verify the model and simulation results, we perform a set
of systematic measurements to characterize the effects of modulation on the beam dynamics  observing directly the island tunes,
the increased coherent damping and the changes to the longitudinal electron beam distribution caused by the modulation.

The paper is organized as follows. We begin (Section II) with the motivation that led us to introduce phase
modulation routinely in user shifts at the LNLS storage ring and with a report of our attempts to stabilize the beam after the installation of the new
RF cavity. In Section III we calculate the electron equations of motion when subjected to phase modulation of the RF field at the second harmonic
of the synchrotron frequency using the hamiltonian formalism and,
using the Fokker-Planck equation, we derive the particle distribution in phase space. Also in Section III we
calculate the dynamics for small oscillations around the stable fixed points created when modulation is turned on and,
with the standard formalism used to calculate Landau damping, we find the increase in
the amount of damping of centroid oscillations related to phase modulation. In this section we also calculate the
longitudinal beam transfer function and the stability diagram showing that the stable
area is increased when phase modulation is turned on. In Section IV
we present results of simulations of the effects of phase modulation in single bunch and multibunch modes. In those simulations one
of the circulating bunches has an internal structure whereas all others are treated as macroparticles. Finally, in Section V, we compare the
results of theory and simulation with experiments in which we measure the island structure, the frequency
for small oscillations around the stable fixed points created by RF phase modulation, the increase in the amount of damping
due to modulation and the longitudinal beam transfer function (BTF), verifying the results obtained
with the theoretical model presented in Section III. Conclusions are given in Section VI.

\section{CBM STABILIZATION BY PHASE MODULATION}

The LNLS synchrotron light source is based on a \mbox{1.37 GeV} electron storage ring and operates with
250 mA of initial current in normal users shifts. A summary of the main parameters for the
Brazilian machine is shown in Table \ref{TabParametrosLNLS}.
At the end of 2003 the RF system of the ring was upgraded and a new cavity was installed \cite{UpgradedLNLS,CommissioningLNLS}.
This second RF cavity was added to the system in order to prepare it for power demands
that the future installation of insertion devices will bring about. After operation with two cavities
was initiated, the beam suffered from frequent instability
outbreaks during user runs.

\begin{table}
\begin{center}
\caption{LNLS ring main parameters.} \vspace{0.2cm}
\begin{tabular*}{7.5cm}{p{4.0cm}cc}
\hline\hline Parameter& Symbol &Value  \\
\hline Beam Energy& E & 1.37 GeV\\
Natural energy spread& $\sigma_{\epsilon}/E$& $5.4\times 10^{-4}$\\
Circumference & C& 93.252 m\\
RF frequency &f$_{rf}$& 476.066 MHz\\
Harmonic number&h & 148\\
Momen. comp. factor &$\alpha$& $8.3\times 10^{-4}$\\
Radiation loss per turn &U$_{0}$& 114 keV \\
Synchronous Phase$^{\ddag}$ & $\phi_{s}$& 166.8$^{o}$\\
Synchrotron Frequency$^{\ddag}$ & $f{s}$& 26 kHz\\
\hline\hline \multicolumn{3}{l}{\footnotesize{$\ddag$ Calculated
using V$_{rf}$=500 kV.}}
\end{tabular*}
\label{TabParametrosLNLS}
\end{center}
\end{table}

\begin{figure}[htb]
\centering
\includegraphics*[width=85mm]{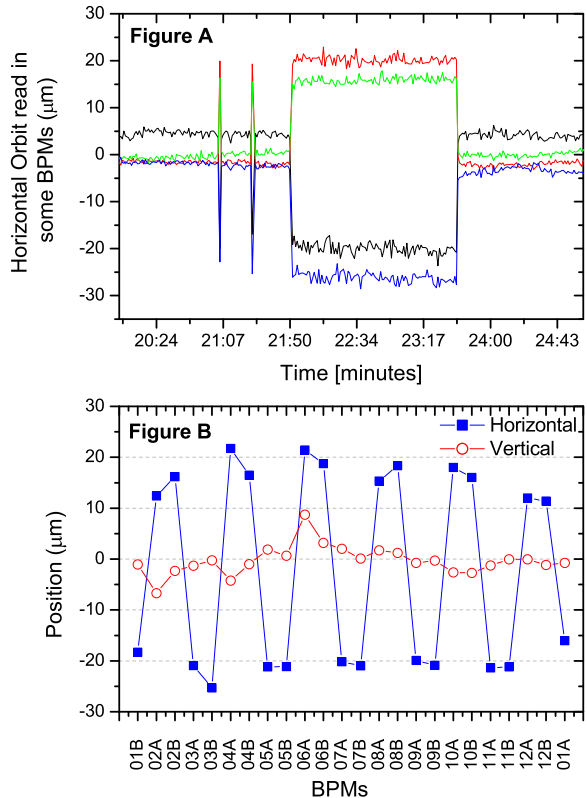}
\caption{Orbit distortion caused by a longitudinal CBM. The
distortion occurs only in the horizontal orbit and has the same
6-fold symmetry of the machine. (A) Time evolution of the horizontal orbit measured by four different BPMs during a user shift
showing the orbit fluctuation caused by the instability excited by a HOM.
(B) Profile of the orbit fluctuation along the storage ring. This is obtained
by subtracting the orbit right before the fluctuation from the orbit right after it. } \label{fig-1}
\end{figure}

The typical signature of the beam instability is an orbit fluctuation measured only in the
horizontal plane, as shown in Figure~\ref{fig-1}. The distortion
pattern has the same 6-fold symmetry of the storage ring and shows
up in the form of a fast transition between two states. For most
of the user run periods, small orbit fluctuations with amplitudes of
$\pm 3\: \mu$m were observed and could be detected at the most
sensitive beamlines. However, larger amplitude distortions were
sometimes observed as shown in Figure~\ref{fig-1}.

The origin of the orbit fluctuations was identified as being
related to large amplitude dipolar longitudinal oscillations
caused by the interaction of the beam with HOMs excited in the
cavities. In fact, the second order term of the non-linear
dispersion function $\eta_{1}(s)$ was measured and has the same
profile observed in the beam position monitors (BPM's) reading and shown in Figure~\ref{fig-1}B.
When a longitudinal dipolar oscillation is
present due to the influence of
a HOM, the relative energy deviation is given by $\delta (t) =
\delta_0 \cos{\Omega_s t}$. The closed orbit distortion is then

\begin{equation}\label{eq:COD}
x_\epsilon (s,t)={\delta (t) \eta (s) + \delta^2 (t) \eta_1 (s)}.
\end{equation}

Since the oscillation is fast (the synchrotron frequency
$\Omega_s$ is of the order of $25\,\hbox{kHz}$) the BPMs measure
only average positions

\begin{equation}\label{eq:averageCOD}
\langle x_\epsilon \rangle = {1 \over 2} \eta_1 (s) \delta_0^2,
\end{equation}

\noindent whereas the maximum amplitude of the oscillations is given by

\begin{equation}\label{eq:CODmax}
x_\epsilon^{max} (s) =\delta_0 \eta (s) =\sqrt{\frac{2 \langle
x_\epsilon \rangle}{\eta_1 (s)}} \eta (s)\mbox{.}
\end{equation}

\noindent $x_\epsilon^{max}$ shows up in the detectors of the beam
lines as an effective increase of the horizontal beam size.
The same orbit distortion can be obtained by phase modulating (at the synchrotron frequency) the
RF master generator that drives the RF stations.

Attempts to eliminate the instabilities by changing the parameters of both
cavities have proved to be not very effective. The search for a
convenient operational point of the cavity was performed
by scanning the parameters that can be used to control the cavity
tuning (temperature, plunger position and axial deformation) and measuring the corresponding HOM frequencies.

\begin{figure}[htb]
\centering
\includegraphics*[width=85mm]{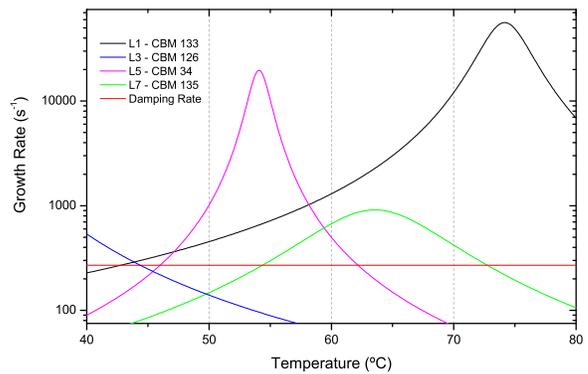}
\caption{CBM growth rates for the new RF cavity in the best
operation point for the plunger position, calculated for
$E_0=1.37\:\hbox{GeV}$ and $I=250\,\hbox{mA}$. Notice the absence
of a mode free region.} \label{fig-2}
\end{figure}

The identification of the potentially harmful modes was
possible by comparing them with the modes measured for a similar cavity used in our injector booster
synchrotron, which had been fully characterized in the laboratory, and by
comparing the frequency shifts of the modes with those predicted
from their simulated field maps. The procedure adopted to analyze the effect of the HOMs is the same
used for similar cavities operating at ELETTRA and
ANKA \cite{Cavidade2} and consisted in the determination of the
growth rates of the coupled bunch mode (CBM) instabilities as a function of
temperature and plunger positions. By mapping the variation of the
modes with these parameters, we determine the
regions for which the growth rates of the oscillations are below the radiation damping
rate, which for the LNLS storage ring at nominal operating energy is $1/\tau_{rad}\approx 270$ s$^{-1}$, as shown in Figure \ref{fig-2}.

The results of the HOM survey showed a comfortable situation for
the old cavity. The temperature range for which there are no
dangerous longitudinal HOMs is quite wide. Unfortunately, the same
is not true for the new cavity. A list of some of its main
longitudinal modes is shown in Table~\ref{tupkf001-t1}. $T_{C}$ is
the critical temperature of the mode, defined as the temperature
at which the frequency of the HOM matches the frequency of the CBM
oscillation. The values of $Q_L$ have been measured in the new
cavity but those of $R_{s}/Q_{0}$ are simulated values. The main problem
with the new cavity is the longitudinal mode L1 whose critical
temperature is just at the edge of the available temperature control range
but is rather insensitive to changes of the cavity tuning parameters, as shown in Figure \ref{fig-2}.
An attempt was made to shift that mode using the plunger
already installed in the cavity but the extent of the plunger
motion was not enough to accomplish the task \cite{CommissioningLNLS}.

\begin{table}[hbt]
\begin{center}
\caption{Main TM$_{0mn}$ Modes of Cavity 2}
\begin{tabular}{lccccc}
\hline\hline \textbf{Mode} & \textbf{f} & \textbf{$R_{s}/Q_{0}$} &
\textbf{$Q_{L}$} & \textbf{$T_{C}$} & CBM
\\
  & (MHz) & $(\Omega)$ &  & $(^{\circ}C)$ & n
\\ \hline
L1 & 904.128 & 29.7 & 20000 & 74 & 133 \\
L3 & 1356.89 & 5.4 & 4000 & 31 & 126 \\
L5 & 1538.25 & 9.5 & 4000 & 54 & 34 \\
L9 & 2040.125 & 8.2 & 25000 & 28 & 43 \\ \hline\hline
\end{tabular}
\label{tupkf001-t1}
\end{center}
\end{table}

\begin{figure}[htb]
\centering
\includegraphics*[width=85mm]{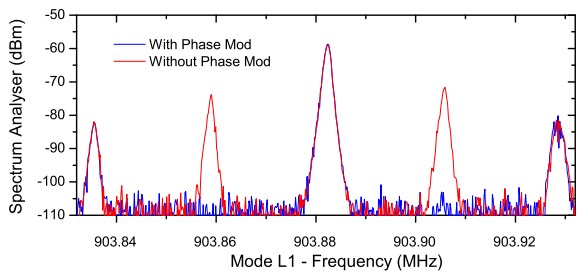}
\caption{Measured frequency spectrum of the signal from a button pick-up in the storage ring demonstrating
the suppression of dipolar synchrotron sidebands as a result of RF phase modulation at the second harmonic of the synchrotron frequency.} \label{fig-3}
\end{figure}

The coupled bunch mode associated
with the L1 mode, CBM $\sharp133$, is always present in the beam
spectrum except in those regions where the growth rate of other
modes, such as CBM $\sharp126$ associated to mode L3, are larger.
HOM mapping was useful for the identification of the best
operation point but this point turned out to be not good enough. As a result, an active
solution in the form of phase modulation of the RF fields at approximately twice
the synchrotron frequency was attempted with remarkable success. Figure \ref{fig-3} shows the longitudinal spectrum of the stored beam
as measured by a button pick-up. The suppression of the dipolar synchrotron sideband of the harmonic
\#281 of the revolution frequency by more than 40 dB illustrates the damping of this CBM by RF phase modulation.

\section{THEORETICAL MODEL}

\subsection{Longitudinal Hamiltonian with Phase Modulation}

We initially follow the standard hamiltonian analysis for the
longitudinal dynamics with phase modulation \cite{Orsini} in which we expand the complete Hamiltonian near the second harmonic resonance
of the synchrotron frequency and retain only the lower order terms. The Hamiltonian for
an electron in a bunch subjected to phase modulation can be
written as follows
\begin{widetext}
\begin{eqnarray}
H(\delta,\phi;t)&=&\frac{\omega_{s}\delta^{2}}{2}+\omega_{s}\tan\bar{\phi}[\sin\phi\cos(A_{m}sin\omega_{m}t)+\cos\phi\sin(A_{m}sin\omega_{m}t)]\nonumber\\
&-&\omega_{s}\cos\phi\cos(A_{m}sin\omega_{m}t)+\omega_{s}\sin\phi\sin(A_{m}sin\omega_{m}t)-\omega_{s}\tan\bar{\phi_{s}}
\label{HamiltonianoTotal}\end{eqnarray}
\end{widetext}
where $\omega_{s}=2\pi f_{s}$ is the synchrotron frequency,
\mbox{$\bar{\phi_{s}}=\pi - \phi_{s}$} where $\phi_{s}$ is the synchronous
phase, $A_{m}$ is the modulation amplitude and $\omega_{m}$ is the
modulation frequency. Using the action-angle variables $(J,\Psi)$
we make the following transformation
\begin{subequations}
\begin{eqnarray}
\delta&=&-\sqrt{2J}\sin(\psi+\omega_{m}t/2+\pi/4), \\
\phi&=&\sqrt{2J}\cos(\psi+\omega_{m}t/2+\pi/4),
\end{eqnarray}
\end{subequations}
which describes the trajectories of the particles in the phase space near the second harmonic resonance in
a reference frame which rotates at half of the modulation frequency.
The resulting time averaged Hamiltonian $K$ [Appendix \ref{ApendiceA}] is then:
\begin{equation}
\langle K\rangle_{t}=\left(\omega_{s}-\frac{\omega_{m}}{2}
\right)J-\frac{\omega_{s}J^{2}}{16}+\frac{\omega_{s}\epsilon J}{4}\cos2\psi
\label{hamiltoniano}\end{equation} where
$\epsilon=A_{m}\tan\bar{\phi}_{s}$

The fixed points of the system are the solutions of
\begin{eqnarray}
\frac{d J}{dt}=-\frac{\partial K}{\partial \psi}=0
\quad \mbox{and}  \quad \frac{d \psi}{dt}=\frac{\partial
K}{\partial J}=0
\end{eqnarray}
and are
\begin{eqnarray}
J_{SFP}=\left\{ \begin{array}{cl}
8\left(1-\frac{\omega_{m}}{2\omega_{s}}\right)+2\epsilon &,\quad
\omega_{m}\leq (2 + \epsilon/2) \omega_{s}\\
0&,\quad \omega_{m} > (2 + \epsilon/2) \omega_{s}
\label{eq:JSFP}\end{array}\right.
\end{eqnarray}
corresponding to the angles $\psi_{SFP}=0, \pi$, which are stable
since $\cos2\psi>0$, and the unstable fixes points are
\begin{eqnarray}
J_{UFP}=\left\{ \begin{array}{cl}
8\left(1-\frac{\omega_{m}}{2\omega_{s}}\right)-2\epsilon & ,\quad
\omega_{m}\leq (2 - \epsilon/2) \omega_{s}\\
0 &,\quad \omega_{m} > (2 - \epsilon/2) \omega_{s}
\end{array}\right.
\end{eqnarray}
corresponding to the angles $\psi_{UFP}=\pi/2, 3\pi/2$.

The phase space coordinates of the new stable fixed points are related to the square root of $J_{SFP}$ or $J_{UFP}$
and are functions of the RF phase modulation amplitude and frequency.

\begin{figure}
\begin{center}
  \includegraphics[width=6cm]{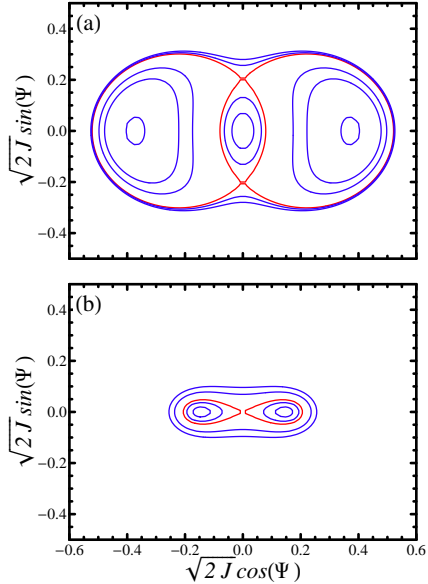}\\
  \caption{Phase space of the modulated bunch with two different modulation frequencies in the rotating reference frame.
  (a) $\omega_{m}\leq (2 -\epsilon/2) \omega_{s}$ with 3 stable fixed points and (b) $\omega_{m} > (2 - \epsilon/2) \omega_{s}$
  with 2 stable fixed points. The red line corresponds to the separatrix.} \label{EFMod}
\end{center}
\end{figure}

The effect of phase modulation is to create new regions of
stability inside the bunch, besides the original one. As the
formation of those islands depends on the
modulation amplitude and frequency, there is a continuum of forms from one island to the
appearance of three islands when \mbox{$\omega_{m}\leq (2 -
\epsilon/2) \omega_{s}$} and the formation of only two islands when
\mbox{$\omega_{m} > (2 - \epsilon/2) \omega_{s}$}. The longitudinal phase
space for the situations described above is shown in Figure \ref{EFMod}.

\subsection{The Effect of Radiation}

All the arguments given above do not take into account the fact that
electrons in a bunch lose energy through the emission of photons.
This effect causes oscillations inside the bunch to damp, so it is
important to know when this damping can destroy the formation
of islands produced by phase modulation or modify some characteristics derived using the hamiltonian formalism.
The analysis that follows uses a simple non-linear oscillation theory to calculate the
trajectories of a single particle under the influence of phase modulation at the second harmonic of the synchrotron
frequency.

The equation of motion for a particle subjected to phase modulation can be written as follows \cite{Sakanaka-damp}
\begin{eqnarray}
\frac{d^{2}\phi}{dt^{2}}+2\gamma_{d}\frac{d\phi}{dt}+\omega_{s}^{2}[1+\epsilon \cos(\omega_{m}t)]=\beta\phi^{3}
\label{eqmotion_damp}\end{eqnarray}
with $\gamma_{d}$ the radiation damping time and
\begin{eqnarray}
\epsilon=A_{m}\tan\bar{\phi}_{s} \quad \mbox{and}\quad \beta=\frac{1}{6}\omega_{s}^{2}.
\end{eqnarray}
We also consider that $A_{m}$, the phase modulation amplitude, and $\phi$ are such that $A_{m}\ll\phi\ll 1$ rad and
retain only the terms most important which are the parametric resonance term on the left side and the third order term on the right side of equation
(\ref{eqmotion_damp}). Non-resonant terms in $A_{m}$ and $\tan\bar{\phi}_{s}$ are discarded.

From the standard analysis of parametric resonances  \cite{Landau} we expect that the solution of equation (\ref{eqmotion_damp}) has the form
\begin{equation}
\phi(t)=a(t)\cos\left(\frac{\omega_{m}}{2}t\right)+b(t)\sin\left(\frac{\omega_{m}}{2}t\right).
\label{eq:ansatzradiation}\end{equation}
Substituting \ref{eq:ansatzradiation} in equation \ref{eqmotion_damp}, we can write a set of equations for the functions
$a$ and $b$
\begin{eqnarray}
2\dot{a}+2\gamma_{d}a+(\alpha+\delta\omega)b+kb(a^{2}+b^{2})=0,\label{eq:saturation1}\\
2\dot{b}+2\gamma_{d}b+(\alpha-\delta\omega)b-ka(a^{2}+b^{2})=0,
\label{eq:saturation2}\end{eqnarray}
where
\begin{equation}
\alpha=\frac{\omega_{s}\epsilon}{2}, \quad k=\frac{\omega_{s}}{8}, \quad \delta\omega=\omega_{m}-2\omega_{s}.
\end{equation}
If a particle is to be captured by the parametric resonance, the slowly varying coeficientes $a(t)$ and $b(t)$ must
increase exponentially. Taking only the lower terms in $a$ and $b$ in
equations (\ref{eq:saturation1}) and (\ref{eq:saturation2}) and using this condition we find that, in order for phase modulation
to create stable islands in the longitudinal phase space, it is necessary that
the condition $\alpha > 2 \gamma_{d}$ be satisfied. Rewriting this condition in terms of the phase modulation amplitude we find
that\footnote{Calculated for a total accelerating voltage of 500 kV.}
\begin{equation}
A_{m}\geq\frac{4\gamma_{d}}{|\tan\phi_{s}|\omega_{s}}\approx 0.026 \, \mbox{rad} \label{raditation_limit}
\end{equation}
and this is a limit imposed by radiation damping bellow which no island is observed.

Changing the longitudinal phase space reference frame to one which rotates with a frequency of $\omega_{m}/2$ in
time we find that the coordinates in this new frame are related only to the functions $a(t)$ and $b(t)$ and the fixed points
are given by the solution of equations (\ref{eq:saturation1}) and (\ref{eq:saturation2}) when $\dot{a}=\dot{b}=0$. Rewriting $a=\sqrt{2r}\cos\theta$ and
$b=\sqrt{2r}\sin\theta$ we end up with the following solutions
\begin{eqnarray}
r=\frac{-\delta\omega\pm\sqrt{\alpha^{2}-4\gamma_{d}^{2}}}{2k},\label{eq:Ampliefetiva}\\
\tan\theta=\frac{-\alpha\pm\sqrt{\alpha^{2}-4\gamma_{d}^{2}}}{2\gamma_{d}}.
\end{eqnarray}

If we set $\gamma_{d}=0$ we find that the stable fixed points are the ones already calculated in the previous section using the hamiltonian
formalism. To take into account the effect of radiation damping we define, using (\ref{eq:Ampliefetiva}), an effective amplitude of modulation which is modified by the radiation damping effect as:
\begin{equation}
A_{m}^{eff}=\sqrt{A_{m}^{2}- \left( \frac{4\gamma_{d}}{|tan\phi_{s}|\omega_{s}} \right)^{2}}.
\end{equation}
This quantity is particulary important when one uses the hamiltonian formalism to make
predictions about measurements. In these calculations $A_{m}^{eff}$ should be used instead of $A_{m}$ when the damping effects are not
negligible.


\subsection{Dynamics of Small Oscillations Around the Fixed Points}

We now consider the dynamics of particle oscillations in the vicinity of the stable fixed points (SFPs)
in order to derive two important properties of the electron motion: the equilibrium density
distribution close to the SFPs and the frequency of synchrotron oscillations around those points (also known as the island tune),
including its dependence on the oscillation amplitude. This is
an important quantity since it describes the frequency of incoherent
oscillation of the particles inside the island when subjected to a
longitudinal kick and its dependence on oscillation amplitude is intimately related to the effectiveness of phase modulation as a damping mechanism
for coherent synchrotron oscillations.

We start by expanding equation (\ref{hamiltoniano}) around a fixed point with coordinates $(J_{0},\psi_{0})$ in the rotating reference frame \cite{Mod1-5}
\begin{subequations}
\begin{eqnarray}
\delta'&=&-\sqrt{2J}\sin \psi+\sqrt{2J_{0}}\sin \psi_{0},\\
\phi'&=&\sqrt{2J}\cos \psi-\sqrt{2J_{0}}\cos \psi_{0},
\end{eqnarray}
\end{subequations}
and find that
\begin{equation}
H'=\frac{A}{2}\delta'^{2}+\frac{B}{2}\phi'^{2}+ \mathcal{O}(\phi'^{3})\label{Hexpand1}
\end{equation}
where
\begin{eqnarray}
A=\omega_{s}-\frac{\omega_{m}}{2}-\frac{\omega_{s}J_{0}}{8}-\frac{\omega_{s}J_{0}}{4}\sin^{2}\psi_{0}\nonumber\\
-\frac{\omega_{s}\epsilon}{4}\left(1-\frac{J_{0}}{6}+\frac{J_{0}}{4}\cos 2\psi_{0}\right),\\
B=\omega_{s}-\frac{\omega_{m}}{2}-\frac{\omega_{s}J_{0}}{8}-\frac{\omega_{s}J_{0}}{4}\cos^{2}\psi_{0}\nonumber\\
+\frac{\omega_{s}\epsilon}{4}\left(1+\frac{J_{0}}{6}+\frac{J_{0}}{4}\cos
2\psi_{0}\right).
\end{eqnarray}
In order to find the steady-state phase space profile of the electron beam under phase modulation around a stable fixed point, we must solve the Fokker-Planck equation,
using $H'$ from equation (\ref{Hexpand1}) and taking into account the radiation damping ($\gamma_{d}$) and quantum excitation ($\kappa=\sigma_{\delta}^{2}\gamma_{d}$) terms.
When all forces are balanced the solution of the Fokker-Planck equation is a bi-gaussian where
\begin{equation}
\Psi(\delta',\phi')=\frac{1}{2\pi
\sigma_{\delta}\sigma_{\phi}}exp\left(-\frac{1}{2\sigma_{\delta}^{2}}\delta'^{2}-\frac{1}{2\sigma_{\phi}^{2}}\phi'^{2}\right)
\label{distribfinal}\end{equation}
and the rms bunch energy and phase spread are
\begin{equation}
\sigma_{\delta}=\sqrt{\frac{\kappa}{\gamma_{d}}}
\qquad\mbox{and}\qquad
\sigma_{\phi}=\sqrt{\frac{A}{B}}\;\sigma_{\delta}=\sqrt{\frac{A\kappa}{B\gamma_{d}}}
\label{comprimentos1}\end{equation}
where
\begin{equation}
\gamma_{d}=\frac{1}{\tau_{rad}}\approx 250 \: s^{-1} \quad
\mbox{and} \quad \kappa\approx 4
\times 10^{-4} s^{-1}.
\end{equation}
In other words, particles close to the stable fixed points are subjected to an effective harmonic potential
and we estimate the steady-state phase space profile of the electron beam under phase modulation (when radiation excitation and damping
are balanced) by a bi-gaussian distribution.

If we go further in expanding (\ref{hamiltoniano}) we find that
\begin{eqnarray}
H'&=&\frac{A}{2}\delta'^2+\frac{B}{2}\phi'^{2}-\frac{\omega_{s}\sqrt{2J_{SFP}}}{16}\phi'^{3}
\nonumber\\&&-\frac{\omega_{s}}{16}\left(\frac{\phi'^{4}+\delta'^{4}}{4}\right)
-\frac{\omega_{s}\epsilon}{24}\left(\frac{\phi'^{4}-\delta'^{4}}{4}\right)
\end{eqnarray}
and that the unperturbed frequency, or island frequency, is given by $\omega=\sqrt{AB}$. This quantity is real for the stable fixed points and imaginary for the unstable ones.

Changing variables to
\begin{equation}
\phi'=\sqrt{2J'}\sqrt{\frac{A}{B}}\cos\Psi' \quad \mbox{and} \quad
\delta'=-\sqrt{2J'}\sqrt{\frac{B}{A}}\sin\Psi'
\end{equation}
and using the canonical perturbation technique \cite{SYLee} with the generating function
\begin{equation}
F_{2}(\Psi',I)=\Psi'I+G_{3}(I)\sin3\Psi'+G_{1}(I)\sin\Psi',
\end{equation}
where the functions $G_{1}(I)$ and $G_{3}(I)$ are chosen so that the term in $J^{3/2}$ is cancelled,
we find that the phase averaged Hamiltonian is
\begin{eqnarray}
\langle H\rangle_{\Psi}(I) \approx \omega I-\frac{\omega_{s}}{16}I^{2}\left[\frac{3}{8}\frac{A^{2}+B^{2}}{A^{2}B^{2}}
+\frac{\epsilon}{4}\frac{A^{2}-B^{2}}{A^{2}B^{2}}\right]\nonumber\\
-\frac{15}{16}\frac{\omega_{s}^{2}\sqrt{I_{SFP}}}{\omega}\left(\frac{A}{B}\right)^{3/2}I^{2}
\end{eqnarray}
and the revolution frequency  around each stable fixed point for a particle with a phase amplitude $\hat{\phi}$ with respect to that point is
\begin{equation}
\omega(\hat{\phi})\approx \omega\left(1-\frac{3\,\omega_{s}}{16}\,\frac{A^{2}+B^{2}}{A^{2}|B|}\,\frac{\hat{\phi}}{8}\right),
\label{eq:freqisland}\end{equation}
where the parameters of the RF phase modulation are in the coefficients $A$ and $B$.


\subsection{Beam Transfer Function and Landau Damping}

To explain why phase modulation has such an intense effect on the
amplitude of CBM oscillations, we consider that phase modulation creates a spread in the frequency
distribution inside the bunch thus increasing the amount of Landau
damping \cite{chao}. In this case we are only interested in the damping of synchrotron oscillations of the bunch centroid which are produced
by the coupling between the synchrotron motion and a longitudinal HOM of the RF cavity.
We therefore proceed to calculate the response of the beam centroid motion to an external harmonic perturbation when phase modulation is on.

Suppose the particles in a bunch have a distribution
$\Psi(r)$ in phase space. In case the beam is phase modulated the distribution can
be approximated, in the rotating reference frame, by three gaussian functions each one centered at
a fixed point
\begin{eqnarray}
\Psi_{0}(r)=\frac{N_{c}}{\sqrt{2\pi}\sigma_{c}}\,e^{-r^{2}/2\sigma_{c}^{2}}+
\frac{N_{i1}}{\sqrt{2\pi}\sigma_{i}}\,e^{-(r-r_{i1})^{2}/2\sigma_{i}^{2}}\nonumber\\+
\frac{N_{i2}}{\sqrt{2\pi}\sigma_{i}}\,e^{-(r-r_{i2})^{2}/2\sigma_{i}^{2}},
\end{eqnarray}
where the indexes $c$ and $i1,2$ indicate the central island or the side islands formed
when the phase modulation is turned on, $r$ is a radial coordinate in the longitudinal phase space and $N_{j}$ is
the relative number of particles in each island such that $N_{i1}+N_{i2}+N_{c}=1$.

Considering that, when an external perturbation with amplitude $F_{0}$ acts on the beam, a dipolar motion
of the centroid is excited, we can write the electron distribution as
\begin{equation}
\Psi(r,\theta,t)=\Psi_{0}(r)+\Psi_{1}(r)e^{j(\Omega t-\theta)}
\label{eq:electrondistrib}\end{equation}
where $\Psi_{1}$ is a small perturbation. Using (\ref{eq:electrondistrib}) and considering that the equation of motion
for each particle inside the island and near the SFPs can be approximated by
\begin{equation}
\ddot{\tau}+2\gamma_{d}\dot{\tau}+\nu_{c,i}^{2}\tau=F_{0}e^{j\Omega t}
\end{equation}
where $\tau$ is the relative time displacement and $\nu_{c,1}$ the island frequency, we can show that \cite{BTF}
\begin{eqnarray}
\Psi_{1}(r)=\frac{N_{c} F_{0}}{\omega_{c}(\Omega-\omega_{c})}\frac{\partial \Psi_{0c}}{\partial r}+
\frac{N_{i1} F_{0}}{\omega_{i}(\Omega-\omega_{i2})}\frac{\partial \Psi_{0i1}}{\partial r}\nonumber\\+
\frac{N_{i2} F_{0}}{\omega_{c}(\Omega-\omega_{i2})}\frac{\partial \Psi_{0i2}}{\partial r}.
\end{eqnarray}
in which we introduced the damping rate due to radiation defining a complex frequency for each island:
$\omega_{c,i}=\nu_{c,i}- j \gamma_{d}$.

The centroid motion is given by the first moment of the phase space distribution
\begin{eqnarray}
\bar{\tau}(\Omega)&=& \int_{0}^{2\pi}\int_{0}^{\infty}r^{2}\cos \theta
d\theta dr \Psi(r,\theta,t)\nonumber\\
&=&\frac{F_{0}}{2\omega_{c}}\left[N_{c}I_{c}(\Omega)+N_{i}\frac{\omega_{c}}{\omega_{i}}I_{i}(\Omega)\right]e^{j\Omega t}
\label{eq:tauLandau}\end{eqnarray}
where the contribution of the side islands result in the same integral since their frequencies are the same.

\begin{figure}
\begin{center}
  \includegraphics[width=7cm]{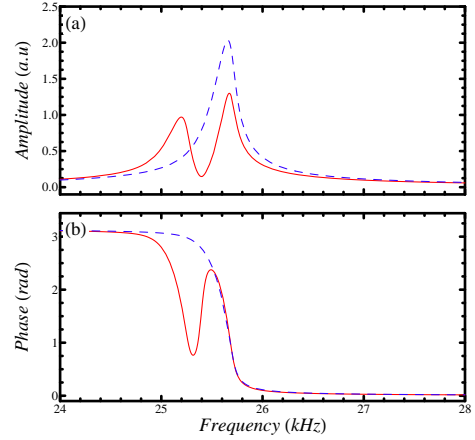}\\
  \caption{Calculated amplitude and phase of the beam transfer function for a bunch with (full red line) and without(dashed blue line)
 RF phase modulation and with half of the particles in the central stable island.}\label{BTF_response}
\end{center}
\end{figure}

The function $I_{n}(\Omega)$ is the dispersion integral and is defined as
\begin{equation}
I_{n}(\Omega)\equiv\pi\int_{0}^{\infty} \frac{r^{2}dr
}{\Omega-\omega_{n}(r)}\frac{\partial \Psi_{0}}{\partial r}
\label{FuncBTF}\end{equation}
and the beam transfer function (BTF) for a beam under phase modulation in the rotating reference
frame is
\begin{equation}
I(\Omega)=N_{c}I_{c}(\Omega)+N_{i}\frac{\omega_{c}}{\omega_{i}}I_{i}(\Omega)
\label{BTFrotationg}\end{equation}
As the measurements are performed in the laboratory reference frame it is necessary to transpose
equation (\ref{BTFrotationg}) to this reference frame. To do this, note that in the lab frame the
frequency of the lateral islands appears as a lower sideband of half of the modulation frequency [Appendix \ref{ApendiceB}] so that we have two different
frequency responses: $\omega_{m}/2+\omega_{c}$ for the central island and $\omega_{m}/2-\omega_{i}$ for the side islands, as shown in Figure \ref{BTF_response}. It is interesting to observe that the effect of phase
modulation is to create a second peak, related to the island frequency and also an intermediary
phase jump in the BTF, these features are very different from the conventional picture expected when measuring the BTF of
an electron beam without phase modulation.

\begin{figure}
\begin{center}
  \includegraphics[width=7cm]{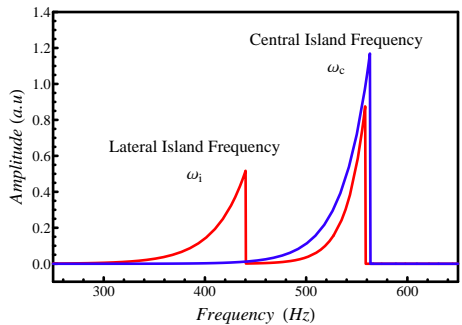}\\
  \caption{Calculated distribution of incoherent frequencies  in the
  rotating reference frame.(blue curve) without phase modulation and (red curve) with phase modulation turned on,
  $f_{m}=50$ kHz, $A_{m}=50$ mrad and half of the particles in the central island.}\label{FreqDistrib}
\end{center}
\end{figure}

It is possible to estimate the effect of phase modulation on the damping
time of the centroid motion if we calculate the increase in frequency
spread due to modulation. The particle distribution in the frequency domain is
given by
\begin{eqnarray}
\Psi'_{0}(\omega)&=&\frac{N_{c}}{\Delta \omega_{c}}\,e^{-(\omega_{c}-\omega)/\Delta \omega_{c}}H(\omega_{c}-\omega)\nonumber\\&&+
\frac{N_{i}}{\Delta \omega_{i}}\,e^{-(\omega_{i}-\omega)/\Delta \omega_{i}}H(\omega_{i}-\omega),
\end{eqnarray}
and shown in Figure \ref{FreqDistrib}, where $H(x)$ is a step function.
Defining the frequency spread as
\begin{equation}
\Delta\omega=\sqrt{\langle \omega^{2}\rangle}=\int\omega^{2}\Psi'_{0}(\omega) \, d\omega \label{eqfreqdispertion}
\end{equation}
we have that the ratio between the natural damping time to the one created when the beam is under phase modulation is
\begin{equation}
\frac{\tau_{mod}}{\tau_{nat}}=\frac{\Delta\omega_{nat}/2\pi+1/\tau_{rad}}{\Delta\omega_{mod}/2\pi+1/\tau_{rad}}.
\label{eqLandauDamping}\end{equation}

The natural spread in the incoherent frequency for the bunch in the
LNLS electron storage ring is \mbox{$(\Delta\omega)^{nat}_{coh}\approx
160$ Hz} and the total damping, including the radiation damping time, related to this frequency dispersion is
\mbox{$\tau_{nat}\approx 3$ ms}. Suppose half of the particles are in the central island, so that $N_{c} = 1/2$,
and the amplitude and frequency of the modulation
are 50 mrad and \mbox{50 kHz} respectively. The frequency spread and the new damping time due to phase
modulation, given by equations (\ref{eqfreqdispertion}) and (\ref{eqLandauDamping}), are \mbox{$\Delta\omega_{mod}=490$ Hz}
and \mbox{$\tau_{mod}=2.2$ ms}. This damping time is smaller than the
estimated growth time for the L1 instability, which is 3.3 ms
\footnote{This value for the mode growth time is for a temperature of \mbox{40 $^{o}$C}, an accelerating voltage of \mbox{500 kV} and a total stored current of \mbox{250 mA}.}.
So, using the phase modulation we are able to set the beam well below the stability boundary and effectively damp this coupled
bunch mode.

So far we have only estimated the effects of phase modulation on beam stability.
We are going now to calculate the implications of the increase in frequency spread due to modulation
on the beam when it is subjected to its own fields.

The spread in incoherent frequencies in the bunch causes it to strongly damp coherent oscillations excited by a harmonic
driving force. This force can be an external excitation or wake field which interacts with the electron beam \cite{chao}.
If we consider a short bunch which interacts with an impedance of the machine, such as a HOM, the centroid motion
is governed by the following equation
\begin{equation}
\frac{d^{2}\tau}{ds^{2}}+\frac{\omega_{0}^{2}}{v^{2}}\tau=-\frac{je^{2}N\eta\omega_{rev}\bar{\tau}}{\beta^{2}E_{0}C^{2}}\mathcal{Z}_{\parallel}
\label{eqcentroide}\end{equation}
where $s=vt$ with $v=\beta c$ is the particle velocity, $N$ is
the number of electrons in the bunch, $e$ is electron charge,
$\omega_{rev}$ is the revolution frequency, $E_{0}$ is the particles energy, $C$ is the orbit length and
$\mathcal{Z}_{\parallel}$ is the effective longitudinal impedance \cite{chao}.

Introducing the ansatz
\begin{equation}
\bar{\tau}(s)=Be^{j\Omega s/v},\label{ansatz}
\end{equation}
the coherent synchrotron tune
shift can be defined as
\begin{equation}
(\Delta\omega_{s})_{coh}=\Omega-\omega_{0}=-\frac{je^{2}N\eta\omega_{rev}c^{2}}{2\omega_{0}
E_{0}C^{2}}\mathcal{Z}_{\parallel}\label{Domega}
\end{equation}
and the stability condition is so that: $Im(\Omega)<0$.

Therefore, using equations (\ref{eq:tauLandau}) and (\ref{Domega}) we can
define the stability diagram for coherent centroid oscillations in the
following way
\begin{equation}
U+iV=-j(\Delta\omega_{s})_{coh}=\frac{j}{I(\Omega)}
=\frac{\hat{g}(\Omega)+j \hat{f}(\Omega)}{\hat{g}^{2}(\Omega)+\hat{f}^{2}(\Omega)}
\end{equation}
where $\hat{f}(\omega)$ and $\hat{g}(\omega)$ are the real and imaginary parts of the BTF $I(\Omega)$ (equation \ref{FuncBTF}).

\begin{figure}
\begin{center}
  \includegraphics[width=7cm]{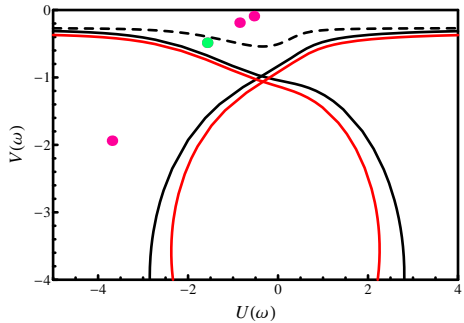}\\
  \caption{Stability diagram in the U x V plane.(black full line) phase modulation on and
   for 50 \% of electrons in the innermost stable island, (blue full line) unstable line for a positive imaginary coherent
   frequency shift, (dashed line) phase modulation off and (dots) various operating points for temperatures of the new cavity varying form 40 to 55 $^{o}$C.}\label{BTF_diagram}
\end{center}
\end{figure}

Figure \ref{BTF_diagram} shows the stability diagram for a bunch under phase modulation
(black full line) and in normal conditions (dashed line). Note that, when phase modulation is turned
on, the stable area is increased and that a new region, inside the loop, is formed.
This new region is also unstable, as shown by the red line which is calculated for positive imaginary coherent
frequency shift, what violates the stability criterion. The dots shown in Figure \ref{BTF_diagram}
represent various operation points depending on the operating temperature of the new cavity. The green mark, for example,
indicates the point for 50 $^{o}$C that is unstable under normal operation conditions and becomes stable if phase
modulation is turned on.

\section{SIMULATION}

In order to gain some insight into the behavior of the bunches under phase modulation
and verify the properties derived from the theoretical model outlined above,
we developed a longitudinal dynamics simulation code including beam loading effects and
the presence of the longitudinal HOMs as well as phase modulation. The tracking code
calculates the trajectories of individual particles in longitudinal phase space and derives
the corresponding averaged quantities representing the motion of the bunch centroid as well as a projected
(time averaged) longitudinal charge density profile appropriate for comparison with experimental results.

Even though the application of the code to the single bunch case is relatively straightforward, the calculations
for the multibunch case are much more time consuming, since at the LNLS storage ring we routinely operate with all 148
bunches filled (no gaps in the bunch train). In order to circumvent that difficulty, we use a \textit{hybrid} simulation,
in which we treat the first 147 bunches ($b_{1}-b_{147}$) as macro-particles (no internal structure) whereas the last bunch
($b_{0}$) is composed of 1000 individual macroparticles. Note that although each one of the 147 bunches without internal
structure is represented basically by a single point in phase space (rather than by a distribution), an effective bunch length
(representing the projection of the phase space bunch distribution on the time axis) is also attributed to those bunches
when calculating the excitation of the wake fields. In fact, as the particles inside $b_{0}$ are redistributed in phase space
as a result of phase modulation and interactions with the environment, its bunch length is also changed and that value of
bunch length is then transferred to all non-structured bunches ($b_{1}-b_{147}$). In this way one important aspect of the
internal bunch structure is shared by all bunches. Clearly, this procedure cannot take into account all properties related
to the internal motion of the bunches and describe their effect on the stabilization of the motion. However, we shall
see in the following sections, that such simulations give additional indication that the increase in incoherent frequency
spread inside the bunches is indeed the main stabilizing mechanism provided by phase modulation.


\subsection{Equations of Motion}

The model used in the simulation considers that every possible bunch
is filled with electrons and each one of those bunches, except one, is represented by
a single macroparticle. The bunches pass periodically through the accelerating cavities
and contribute to the cumulative build-up of the induced fields. The effect of the beam loading is also taken
into account in the calculation of the trajectories of the individual electrons of the structured bunch.

In order to calculate the trajectories of these macroparticles in the longitudinal phase space, we
use the following equations:
\begin{eqnarray}
\tau_{b,n}&=&\tau_{b,n-1}-\alpha \delta_{b,n-1}T_{0},\\
\delta_{b,n}&=&(1-2\gamma_{d}T_{0})\delta_{b,n-1}+\frac{(eV_{tot}-U_{0})}{E},
\end{eqnarray}
with \begin{eqnarray} V_{tot}&=&V_{rf}+V_{bl}
\end{eqnarray}
where $\tau$ and $\delta$ are the time displacement and energy deviation of the
macroparticle related to the reference synchronous particle, $b$ is the index of the bunch and $n$ is the number of turns,
T$_{0}$ is the revolution period, V$_{rf}$ is the gap voltage and
V$_{bl}$ is the energy drained by the beam loading effect, which
takes into account the beam loading of the fundamental mode and of the
longitudinal mode (L1). In order to simulate each particle in the structured
bunch ($b_{0}$), we use the same equations and also consider
radiation excitation.

The phase modulation is included in the calculation of the rf
voltage as follows \begin{equation}
V_{rf}=V_{0}\sin(\phi_{s}-\omega_{rf}\tau_{b,n}+A_{m}\sin(\omega_{m}t))
\end{equation}
were V$_{0}$ is the peak voltage, $\phi_{s}$ the synchronous
phase, $\omega_{rf}$ the angular RF frequency, $A_{m}$ the
modulation amplitude and $\omega_{m}$ the modulation frequency
which we set near the synchrotron second harmonic
($\omega_{m}=2\omega_{s}+\delta\omega$).

The beam loading effect is introduced in the tracking simulation
using the following expressions:
\begin{equation}
V_{n}=V_{n-1}e^{-\omega_{res}\Delta t/2Q_{L}+j\omega_{res}\Delta
t}-kq_{0}e^{-\omega_{res}^{2}\sigma_{\tau}^{2}}
\end{equation}
where $\omega_{res}$ is the angular frequency and $k$ is the
loss factor of the mode, $q_{0}$ is the charge per bunch and $\sigma_{\tau}$ is the
rms bunch length.

\subsection{Single Bunch Simulations}

\begin{figure}
\begin{center}
  \includegraphics[width=7cm]{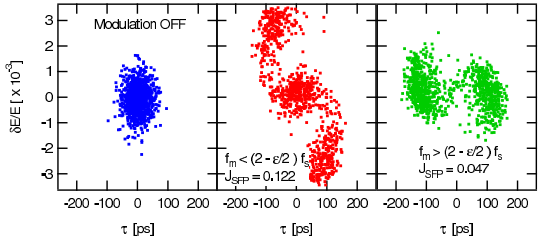}\\
  \caption{Single bunch (1000 macroparticles) simulation showing the phase space for three different situations (blue) without modulation,
  (red) with $f_{m}<(2-\epsilon/2)f_{s}$ and (green)  $f_{m}>(2-\epsilon/2)f_{s}$. In the two last figures it is possible to observe
   the formation of three and two islands respectively, as predicted by the theory.}\label{Simul_island1}
\end{center}
\end{figure}

Using the simulation code described above, we calculate the
phase space distribution of the particles in a single bunch of 1000 macroparticles under phase
modulation. Figure \ref{Simul_island1} shows the phase space with two
different modulation frequencies and also without phase modulation. The phase space patterns observed are in good agreement with the ones
calculated using the Hamiltonian (\ref{hamiltoniano}) and shown in Figure \ref{EFMod}. In particular,
the position of the islands center given by simulation corresponds to the action amplitudes $J_{SFP}$=0.122 and 0.047 for 3 and 2 islands respectively,
as shown in Figure \ref{Simul_island1}. This is in very good agreement with the theoretical values
0.042 and 0.130 for the cases with 3 and 2 islands, obtained from (\ref{eq:JSFP}) using the effective modulation amplitude ($A_{m}^{eff}$).

\begin{figure}
\begin{center}
  \includegraphics[width=7cm]{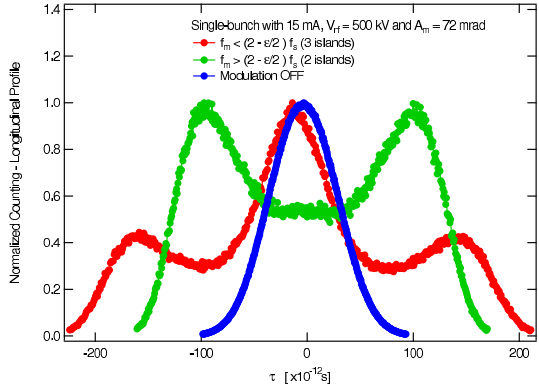}\\
  \caption{Longitudinal profile in time domain of a simulation of 1000 macroparticles in three different situations:
  (blue) without modulation, (red) with $f_{m}<(2-\epsilon/2)f_{s}$ and (green) $f_{m}>(2-\epsilon/2)f_{s}$. Those patterns simulates the measurements
  of the signal of a button pickup.}\label{Simul_island_profile}
\end{center}
\end{figure}

We also calculate the time-averaged longitudinal particle distribution profiles in time domain, which is a histogram
of the particles position in time for a period much longer than the synchrotron one. This result simulates
the profile that would be seen when measuring the signal from a button pickup, and is shown in Figure \ref{Simul_island_profile}.

\begin{figure}
\begin{center}
  \includegraphics[width=7cm]{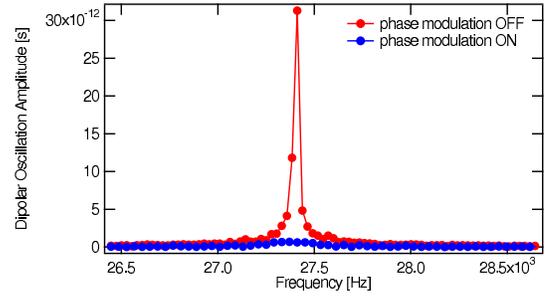}\\
  \caption{Amplitude of the dipolar oscillation of the bunch centroid excited by a RF cavity
  longitudinal HOM for a single bunch with 30 mA.}\label{SBdamping}
\end{center}
\end{figure}

Finally we simulate the damping of a coherent dipolar oscillation by phase modulation. The dipolar oscillation of a high current
single bunch (30 mA) is excited by the L1 HOM and, when phase modulation is turned on, the oscillation amplitude is reduced by a
factor 30 (Figure \ref{SBdamping}), which corresponds approximately to an attenuation of 30 dB. This is similar to the 40 dB damping
of the dipolar spectral line produced by phase modulation (Figure \ref{fig-3}) that we typically observe in multibunch mode. Note
that in this single bunch simulation the effects of an increased incoherent frequency spread are fully taken into account.

\subsection{Multibunch Simulations}

\begin{figure}
\begin{center}
  \includegraphics[width=7cm]{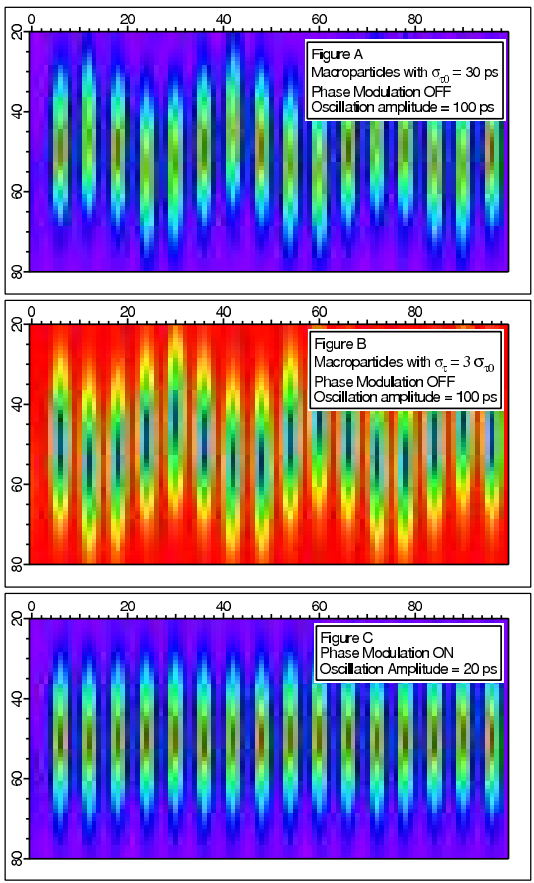}\\
  \caption{Simulation of the centroid oscillation of a longitudinal CBM driven by an RF cavity HOM:
 (A) without phase modulation, (B) without phase modulation and three times the natural bunch length and (C) with phase modulation.The figure shows one of each 6 bunches, as it would be seen by a
   streak camera.}\label{Simul_island_osc}
\end{center}
\end{figure}

For the multibunch case, we simulate the dipolar oscillation excited by the longitudinal HOM of
the cavities (Figure \ref{Simul_island_osc}A). The simulations are performed considering a total stored current of
200 mA. A comparison between two different multibunch tracking
simulations shows that the change in the amplitude of the dipolar
coherent synchrotron oscillation when the L1 mode is turned on is
greater than 100 dB, demonstrating the excitation of dipolar
oscillations in the beam, as observed during users shifts. We could
also observe a good agreement between the value of the growth time
of the L1 mode given by theory (4 ms) and simulation (4.5 $\pm\rm$ 0.5 ms).

Figures \ref{Simul_island_osc}A and B show results of multibunch simulations without phase modulation for two different equilibrium bunch lengths.
We see that even if the bunch length in increased by a factor of 3 (which is larger than the increase observed in multibunch simulations when phase
modulation is applied) there is basically no damping of the coherent oscillations excited by the L1 HOM. In other words,
the mere dilution of the beam bunches is not enough to explain the coherent damping efficiency of phase modulation:
the internal rearrangement of longitudinal phase space and corresponding increased frequency spread is a fundamental
ingredient, as could be seen in the single bunch simulations above.

Finally, we simulate a multibunch case with phase modulation, which is shown in Figure \ref{Simul_island_osc}C. Now we observe
a damping of the CBM amplitude by a factor 5, which is much smaller than what is observed in the single bunch simulations
as well as in experiments. This is a result of the \textit{hybrid} character of our simulation, in which  only one out of
148 bunches has an internal structure (and corresponding frequency spread), although all bunches share the common increased bunch length.
The damping of the structured bunch $b_{0}$
weakens the coupling between the electron beam and the HOM reducing the amplitude of the CBM coherent oscillation.

\section{MEASUREMENTS AND RESULTS}

In Sections III and IV we calculate and simulate some characteristics of the longitudinal beam dynamics with RF phase modulation
on the second harmonic of the synchrotron frequency. In this section we perform an experimental verification of
the predictions made. The details of the experimental setup and the results obtained are shown in the
following parts.


\subsection{Experimental Setup}

\begin{figure}
\begin{center}
  \includegraphics[width=7cm]{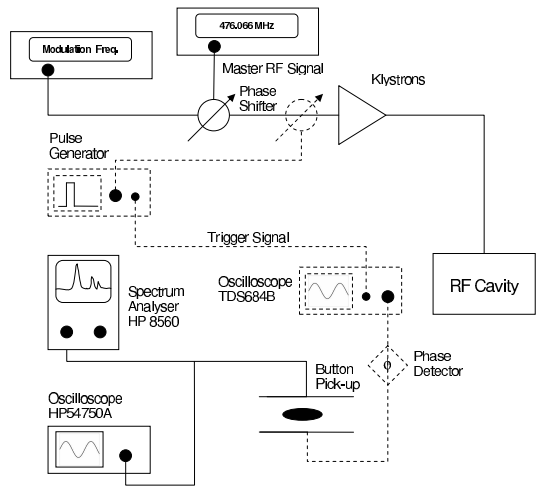}\\
  \caption{General layout of the experimental setup. The solid line correspond to the
  experimental setup used to measure the bunch shape and the beam spectrum while the dashed line
  corresponds to the additional setup used to measure the damping time for coherent synchrotron oscillations.}\label{Meas_Layout}
\end{center}
\end{figure}

The basic idea of the experiments is to excite the beam via phase modulation of the RF fields
and observe its response as a function of a controlled parameter such as modulation amplitude and frequency.
To modulate the phase of the accelerating fields, we use a voltage
driven phase shifter attached to the output of the RF generator
responsible for producing the master RF signal. This phase shifter
is controlled by a signal from another generator that creates
a modulation with a chosen frequency. The full experimental setup
is shown in Figure \ref{Meas_Layout}.

The shape of the bunch is measured by a wideband oscilloscope (HP 54750A, 20 GHz) and the amplitude of the dipolar synchrotron sideband is observed
using a spectrum analyzer (HP8560) both receiving the signal from a button pickup
in the ring. In the multibunch case we choose to observe the synchrotron line around
harmonic $\sharp281$ of the revolution frequency that is close to the frequency of the L1 longitudinal HOM from the
cavity.

We also measure the damping time for synchrotron oscillation of
the centroid of a single bunch by comparing the phase between the master RF
signal to the \mbox{476.066 MHz} component of the longitudinal beam signal and observe the signal in
a high sampling rate oscilloscope (TDS684B). To create a
longitudinal kick we use a function generator that produces a step function with a given
amplitude and acts on a phase shifter in series with the
one used to modulate the RF wave. The additional setup required to measure the damping time is shown by the
dashed lines in Figure \ref{Meas_Layout}.


\subsection{Experimental Results and Comparison with Theory}

In this section we present a comparison between the predictions of the theoretical model and
measurements. This analysis is done in
the following order: island formation as a function of modulation
frequency, damping time of the centroid synchrotron motion as a
function of modulation frequency and amplitude, measurement of island tune and its behavior as a function of
modulation amplitude and frequency, and measurement of the
BTF.

\subsubsection{Island Formation}

\begin{figure}
\begin{center}
  \includegraphics[width=7cm]{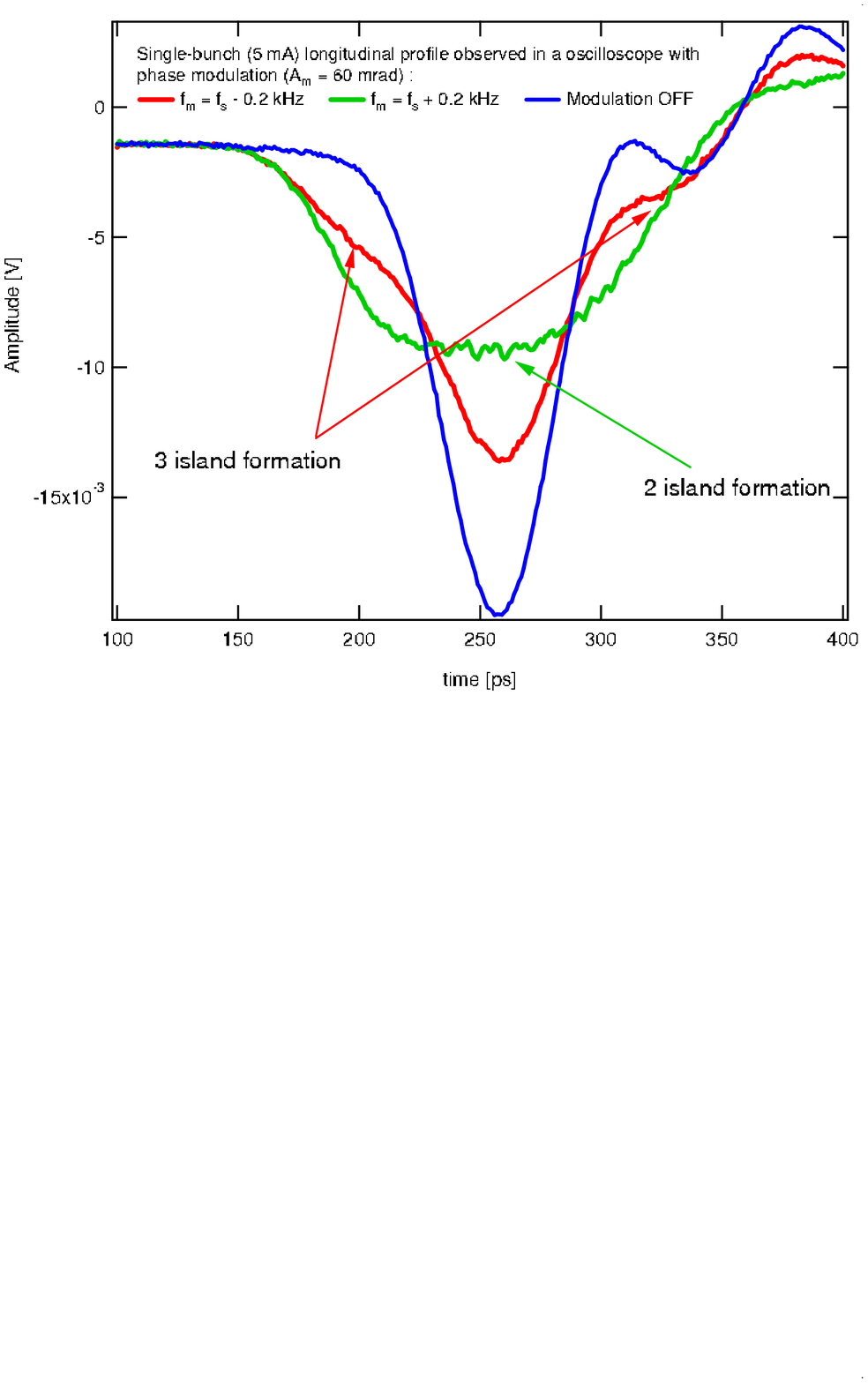}\\
  \caption{Beam profile observed in an oscilloscope for a gap voltage of 460 kV.}\label{Island_Gigao}
\end{center}
\end{figure}

The time-averaged longitudinal bunch profile predicted by our single bunch simulation, Figure \ref{Simul_island_profile}, is compared with the experimental
results of the signal from a button pickup observed on an oscilloscope and shown in Figure \ref{Island_Gigao}. We observe that, for modulation frequencies
$\omega_{m}<\omega_{s}(2-\epsilon/2)$, the profile corresponds to the
case with three stable regions whereas, when
$\omega_{m}>\omega_{s}(2-\epsilon/2)$, only two stable regions are
created. This is an indirect way to observe the bunch internal structure
created by phase modulation. The same result has already been obtained directly in other laboratories \cite{Mod2-1} using
a streak camera to observe the beamlets.

\subsubsection{Longitudinal Damping Time}

\begin{figure}
\begin{center}
  \includegraphics[width=7cm]{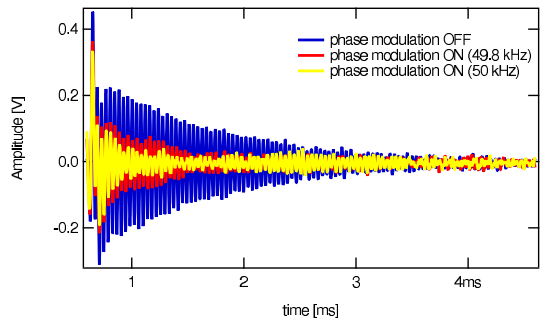}\\
  \caption{Example of a measurement of the synchrotron damping time.}\label{Damping_Example}
\end{center}
\end{figure}

\begin{figure}
\begin{center}
  \includegraphics[width=7cm]{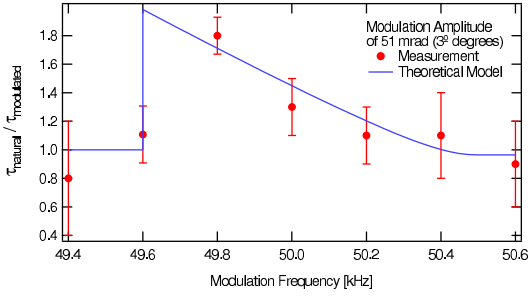}\\
  \caption{Increase in damping time as a function of modulation for a gap voltage of 460 kV:
 (red dots) measurements and (blue curve) theoretical predictions calculated for an island population of 50\%.}\label{IncreaseDamping_Rate}
\end{center}
\end{figure}

The measurements of the damping of coherent synchrotron oscillations are performed
using a function generator and a phase shifter to apply a
longitudinal kick to the beam. Using the signal from a button pick-up, we measure the phase of the RF component of the beam with respect
to the master RF signal. Figure \ref{Damping_Example} is an example of the measurement for
a single bunch with \mbox{4.3 mA} and Figure \ref{IncreaseDamping_Rate} is the damping time ratio for a
single bunch with phase modulation turned on and off.

The results presented in Figure \ref{Damping_Example} are not
the ones observed directly on the oscilloscope. When performing
this measurement we observe that the phase of the RF
component of the beam oscillates not only with the synchrotron frequency but
also with multiples and sub-multiples of the modulation
frequency. The raw data acquired is filtered and
those frequencies removed from the spectrum in order to leave just a band of
frequencies around the dipolar frequency of centroid oscillations, which allows us to determine the damping
time of coherent dipolar oscillations.

A set of measurements of damping time versus modulation frequency is shown in Figure \ref{IncreaseDamping_Rate} and it is
possible to observe that phase modulation is effective only for a
narrow band of frequencies approximately 400 Hz wide. For frequencies
above 50.2 kHz the effect no longer exists, which agrees with the theoretical model
used to describe the damping mechanism. This model predicts that, for frequencies above 50.3 kHz ($f_{m}=(2+\epsilon/2)f_{s}$), the islands
collapse into one so that there is no increase in damping time.

Using equation (\ref{eqLandauDamping}), we can estimate the value for the
damping time increase due to phase modulation and Figure \ref{IncreaseDamping_Rate} shows good agreement between theory and experimental data.
The error bars are relatively large due to the difficulties of measuring the damping time.

\subsubsection{Measurement of the Island Tune}

\begin{figure}
\begin{center}
  \includegraphics[width=7cm]{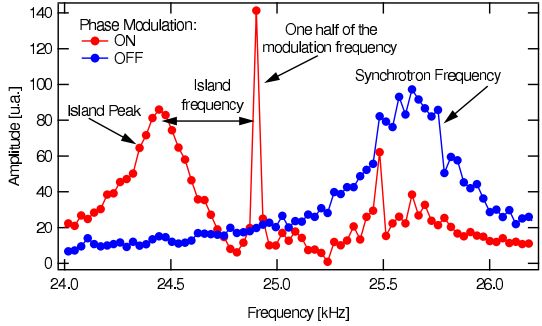}\\
  \caption{Observation of the island tune.}\label{Island_Freq_Example}
\end{center}
\end{figure}

\begin{figure}
\begin{center}
  \includegraphics[width=7cm]{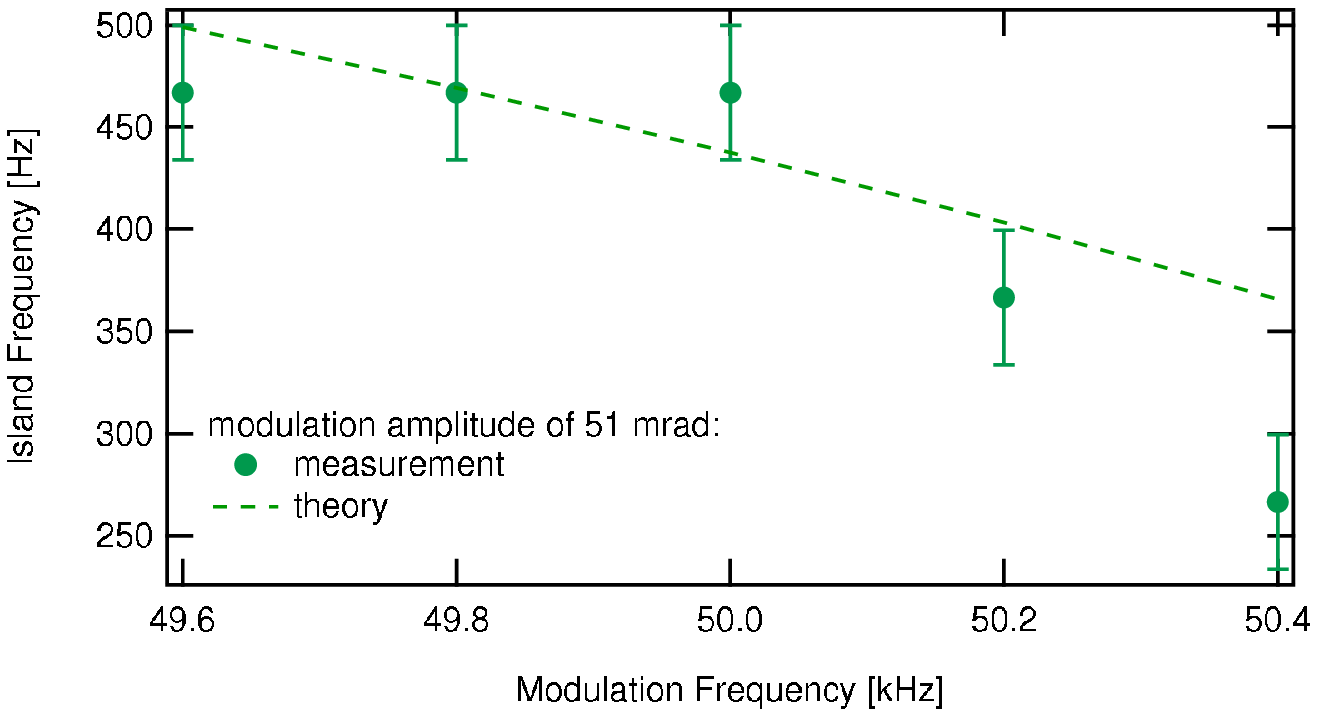}\\
  \caption{Island frequency as a function of modulation frequency for a gap voltage of 460 kV.}\label{IslandFreq_ModFreq}
\end{center}
\end{figure}

\begin{figure}
\begin{center}
  \includegraphics[width=7cm]{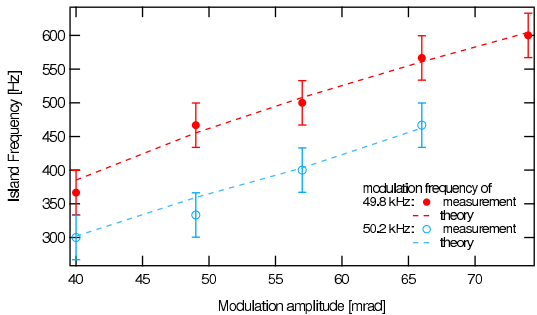}\\
  \caption{Island frequency as a function of modulation amplitude for a gap voltage of 460 kV.}\label{IslandFreq_ModAmpli}
\end{center}
\end{figure}

An FFT of the damping time measurement data also shows a lower sideband of the line
corresponding to half of the modulation frequency. This sideband, shown in Figure \ref{Island_Freq_Example}, correspond to the island tune [Appendix \ref{ApendiceB}].
Figures \ref{IslandFreq_ModFreq} and \ref{IslandFreq_ModAmpli} show the measured dependence of the island frequency
with modulation parameters and the corresponding theoretical curves (\ref{eq:freqisland}), which are in good agreement.

\subsubsection{Measurement of the BTF}

\begin{figure}
\begin{center}
  \includegraphics[width=7cm]{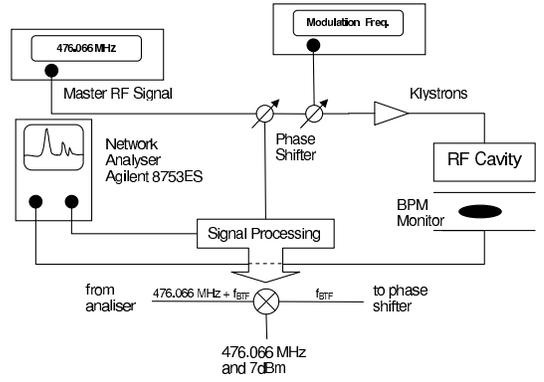}\\
  \caption{Experimental setup used for measuring the BTF.}\label{ExpSetup2}
\end{center}
\end{figure}

The experimental setup used to measure the BTF is shown in Figure \ref{ExpSetup2}. We use a network analyzer to
excite a dipolar oscillation on a low current single bunch and measure the beam response in amplitude and phase using a button pickup.

\begin{figure}
\begin{center}
  \includegraphics[width=7cm]{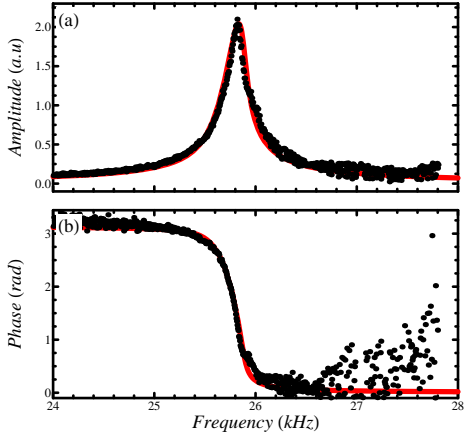}\\
  \caption{Measurement of BTF for a single bunch without phase modulation for a gap voltage of 500 kV.
  The dots correspond to data points and the full red line to theory.}\label{BTFsemmod}
\end{center}
\end{figure}

\begin{figure}
\begin{center}
  \includegraphics[width=7cm]{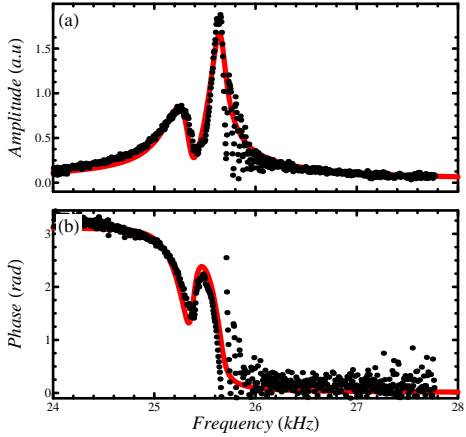}\\
  \caption{Measurement of BTF for a single bunch, for a gap voltage of 500 kV, with phase modulation using the following parameters: $f_{m}=51$ kHz, $A_{m}=$ 35 mrad
  and $N_{c}=$65 \%. The dots correspond to data points and the full red line to theory.}\label{BTFcommod}
\end{center}
\end{figure}

The results for a set of measurements with phase modulation turned on and off are shown in Figures \ref{BTFsemmod} and \ref{BTFcommod} respectively. The dots
are the experimental data and the full line comes from the theoretical model.
The result of those measurements indicates that the phase modulation produces an increase in frequency
spread within the bunch since the width in the amplitude response it related to this spread.

\begin{figure}
\begin{center}
  \includegraphics[width=8cm]{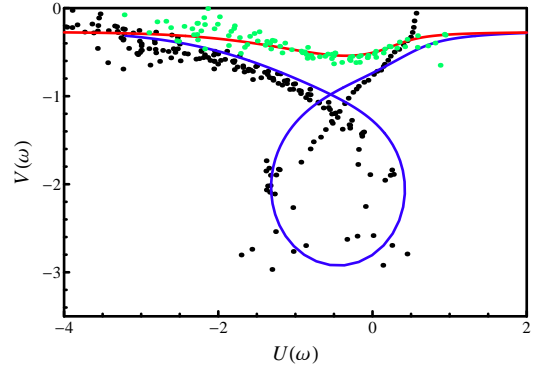}\\
  \caption{Stability diagram of a single bunch with (black dots) and without (green dots) phase modulation and the respective
  theoretical curves.}\label{StabilityDiagram2}
\end{center}
\end{figure}

Figure \ref{StabilityDiagram2} shows the stability diagram for the same set of measurements above
(Figures \ref{BTFsemmod} and \ref{BTFcommod}) and the corresponding theoretical curves. The green dots and red line correspond to the
case without modulation, whereas the black dots and blue line correspond
to a phase modulation amplitude of 35 mrad and frequency of \mbox{51 kHz}.
Note that the stable area is effectively increased when phase modulation is turned on and also that
some features, such as the loop predicted by the model, are present in the experimental data.

\section{DISCUSSION AND CONCLUSIONS}

We have shown experimental and theoretical evidence that, when RF phase modulation on the second harmonic of the synchrotron frequency is used, the
available stable area to the electron beam is effectively increased and that the mechanism responsible for this enhancement in stability
is an increase in the incoherent frequency spread which, via Landau damping, helps to alleviate unstable CBMs.
We also simulate and calculate the island formation and perform measurements that indicate that these structures are indeed formed in the
longitudinal phase space.

Our simulations also indicate that the lengthening of the bunches which results from phase modulation (an effect
which has already been used at several laboratories as a tool to increase Touschek lifetime of intense bunches) cannot
explain by itself the stabilization of CBMs that is observed when phase modulation is applied. In fact, only when bunches with
internal structure (and corresponding frequency spread) are included in the simulations do we observe the intense damping
of CBMs that is obtained in experiments.

Finally, the model developed to describe the BTF measurements presents all interesting features of the experimental results,
namely the double peaks and a second phase jump (which shows up as the loop of the stability diagram), although the
island population still remains as a free parameter. We hope to be able to access this quantity in the future by using a streak camera.

We have been using RF phase modulation since 2004 to stabilize the electron beam of the LNLS storage ring.
Although in the Brazilian machine the overall lifetime increase (we observed a lifetime increase of 30 \%) was not as great as the one observed in other laboratories \cite{Mod4-3,Mod3-2,Mod2-1}
the orbit fluctuation caused by the coupling of the beam with a HOM of the RF cavities was successfully suppressed improving the
beam quality for users.


\subsection*{ACKNOWLEDGEMENT}
We would like to express deep gratitude to
the LNLS RF and diagnostics group whose support
were crucial for the successful completion of this work.
This work was completed with the support of FAPESP (Funda\c c\~ao de Apoio a Pesquisa do Estado de S\~ao Paulo).

\appendix
\section{LONGITUDINAL HAMILTONIAN AROUND THE SECOND HARMONIC RESONANCE}\label{ApendiceA}

Here we present the detailed derivation of the Hamiltonian with phase modulation.
Beginning with \ref{HamiltonianoTotal} and changing variables to
\begin{equation}
\phi=\sqrt{2\tilde{J}}\cos\tilde{\psi} \quad \mbox{and} \quad \delta=-\sqrt{2\tilde{J}}\sin\tilde{\psi},
\end{equation}
we get
\begin{eqnarray}
H(\tilde{J},\tilde{\psi})&=&\omega_{s}\tilde{J}\sin^{2}\tilde{\psi}-\omega_{s}\sqrt{2\tilde{J}}\tan\bar{\phi_{s}}\cos\tilde{\psi} \nonumber\\
&&+\omega_{s}\tan\bar{\phi_{s}}\sin[\sqrt{2\tilde{J}}\cos\tilde{\psi}+A_{m}\sin\omega_{m}t] \nonumber\\
&&-\omega_{s}\cos[\sqrt{2\tilde{J}}\cos\tilde{\psi}+A_{m}\sin\omega_{m}t].
\end{eqnarray}

Expanding the terms in $\sin\Psi$ and $\cos\Psi$ using Bessel functions and restricting our
analysis to even resonances close to the second harmonic of the
synchrotron frequency ($\omega_{m} \approx 2\omega_{s}$), we can simplify the
equation above to
\begin{widetext}
\begin{eqnarray}
H(\tilde{J},\tilde{\psi})&=&\left(\omega_{s}-\frac{\omega_{s}}{2}\cos2\tilde{\psi}\right)\tilde{J}-\frac{\omega_{s}\tilde{J}^{2}}{16}-\omega_{s}\tan\bar{\tilde{\psi}_{s}}\sqrt{2\tilde{J}}\cos\tilde{\psi}+\omega_{s}\tan\bar{\tilde{\psi}_{s}}\sqrt{2\tilde{J}}\sin(\omega_{m}t)\tilde{J}_{0}(\sqrt{2\tilde{J}})
\nonumber\\&&-2\omega_{s}\sum_{k=1}^{\infty}(-1)^{k}\tilde{J}_{2k}(\sqrt{2\tilde{J}})\cos(2k\tilde{\psi})
+2\omega_{s}\tan\bar{\tilde{\psi}_{s}}\sum_{k=0}^{\infty}(-1)^{k}\tilde{J}_{2k+1}(\sqrt(2\tilde{J}))\cos[(2k+1)\tilde{\psi}]
\nonumber\\&&-\omega_{s}\tan\bar{\tilde{\psi}_{s}}A_{m}\tilde{J}_{2}(\sqrt{2\tilde{J}})\sin(\omega_{m}t-2\tilde{\psi}).
\end{eqnarray}
\end{widetext}

The new Hamiltonian is still time dependent and in order to eliminate this
dependence we can perform a canonical transformation using the
following generator function
\begin{equation}
F_{2}(J,\psi)=\left(\tilde{\psi}-\frac{\omega_{m}t}{2}-\frac{\pi}{4}
\right)J
\end{equation}
so that the new coordinates in a rotating reference frame are
\begin{equation}
\tilde{J}=\frac{\partial F_{2}}{\partial\psi}=J \quad\mbox{and}\quad
\psi=\frac{\partial
F_{2}}{\partial J}=\tilde{\psi}-\frac{\omega_{m}t}{2}-\frac{\pi}{4}
\end{equation}
and the Hamiltonian is
\begin{equation}
K=H(J,\psi)+\frac{\partial F_{2}}{\partial
t}=H(J,\psi)-\frac{\omega_{m}}{2}J
\end{equation}
performing a time average, the fast oscillation terms will vanish and we find that
\begin{equation}
\langle K\rangle_{t}=\left(\omega_{s}-\frac{\omega_{m}}{2}
\right)J-\frac{\omega_{s}J^{2}}{16}+\frac{\omega_{s}\epsilon J}{4}\cos2\psi\end{equation}
which is the Hamiltonian around the second harmonic synchrotron resonance.

\section{CALCULATION OF THE ISLAND FREQUENCY IN THE LABORATORY REFERENCE FRAME}\label{ApendiceB}

In the rotating reference frame the motion of the particles in the islands can be
described by
\begin{eqnarray}
\phi'=r_{0}\cos(\omega't)+x_{0}\quad \mbox{and} \quad \delta'=-r_{0}\sin(\omega't)
\end{eqnarray}
where $r_{0}$ is the amplitude of the oscillations around the stable fixed point,
$x_{0}$ is the stable fixed point coordinate in phase space and $\omega'=\sqrt{AB}$ is either the lateral island frequency or the central island frequency.
To go from the rotating frame (which rotates clockwise with respect to the lab frame) to the laboratory reference frame, it is necessary to transform the above set of equations using the
following rotation matrix
\begin{equation}
\mathcal{R}(\omega_{m}t/2)=\left[\begin{array}{cc} \cos(\frac{\omega_{m}t}{2})& \sin(\frac{\omega_{m}t}{2})\\
-\sin(\frac{\omega_{m}t}{2}) &\cos(\frac{\omega_{m}t}{2})\end{array}\right]
\end{equation}
and the new particles coordinates are
\begin{equation}
\left(\begin{array}{c}\phi\\ \delta \end{array}\right)=\mathcal{R}\left(\omega_{m}t/2\right)\left(\begin{array}{c}\phi'\\ \delta'\end{array}\right),
\end{equation}
or writing explicitly for each coordinate
\begin{eqnarray}
\phi&=&r_{0}\cos\left[\left(\omega'+\frac{\omega_{m}}{2}\right)t\right] + x_{0}\cos\left(\frac{\omega_{m}t}{2}\right),\\
\delta&=&r_{0}\sin\left[\left(\omega'+\frac{\omega_{m}}{2}\right)t\right] + x_{0}\sin\left(\frac{\omega_{m}t}{2}\right).
\end{eqnarray}
To understand this set of equations we have first to know wether
particles in the islands rotate clockwise or counterclockwise. Taking the harmonic part of the Hamiltonian (\ref{Hexpand1}) expanded around each stable fixed point we find that
\begin{equation}
\frac{d\phi'}{dt}=A\delta' \quad \mbox{and} \quad \frac{d\delta'}{dt}=B\phi'
\end{equation}
so that if $A>0$ and $B>0$ the particles rotate clockwise and if $A<0$ and
$B<0$ the particles rotate counterclockwise.

Calculating the coefficients $A$ and $B$ for our parameters, we find that
particles in the central island rotates clockwise and particles in the lateral islands rotate counterclockwise.
Using this result we can write the particles equations in the lab reference frame
\begin{eqnarray}
\phi_{i}&=&r_{0}\cos\left[\left(\frac{\omega_{m}}{2}-\omega_{i}\right)t\right] + x_{0}\cos\left(\frac{\omega_{m}t}{2}\right),\\
\delta_{i}&=&r_{0}\sin\left[\left(\frac{\omega_{m}}{2}-\omega_{i}\right)t\right] + x_{0}\sin\left(\frac{\omega_{m}t}{2}\right),
\end{eqnarray}
and
\begin{eqnarray}
\phi_{c}&=&r_{0}\cos\left[\left(\frac{\omega_{m}}{2}+\omega_{c}\right)t\right],\\
\delta_{c}&=&r_{0}\sin\left[\left(\frac{\omega_{m}}{2}+\omega_{c}\right)t\right],
\end{eqnarray}
for the lateral islands and central island respectively.

From the equations above we note that when measuring in the laboratory reference
frame the particles respond to and external excitation with three different frequencies: a narrow line at $\omega_{m}/2$, which is related to the
to the external phase modulation, and a broad line at $\omega_{m}/2-\omega_{i}$ and $\omega_{m}/2+\omega_{c}$ that correspond to the particles incoherent revolution
frequencies around each stable fixed point.
\bibliographystyle{amsplain}

\end{document}